\documentclass[10pt,prl,aps,superscriptaddress,twocolumn,longbibliography]{revtex4-1}
\usepackage{amssymb,amsmath}
\usepackage{graphicx}
\usepackage{color}
\usepackage{hyperref}

\begin{document}

\newcommand{\e}{{\rm e}}
\newcommand{\norm}[1]{\left\lVert#1\right\rVert}
\newcommand{\rmi}{{\rm i}}
\renewcommand{\Im}{\mathop\mathrm{Im}\nolimits}
\newcommand{\red}[1]{{\color{red}#1}}
\newcommand{\blue}[1]{{\color{blue}#1}}
\newcommand{\rot}[1]{\text{rot}\, #1}
\newcommand{\df}[2]{\frac{\partial #1}{\partial #2}}
\newcommand{\bas}[1]{\hat{{\bf #1}}}
\renewcommand{\cite}[1]{[\onlinecite{#1}]}

\newcommand{\ket}[1]{\left|#1\right>}      %Dirac designations
\newcommand{\bra}[1]{\left<#1\right|}
\newcommand{\eps}{\varepsilon}      %Greek epsilon
\newcommand{\om}{\omega}      %Greek omega
\newcommand{\kap}{\varkappa}      %Greek kappa
\newcommand{\VB}[1]{\mathbf{#1}} % Bold font vector

\newcommand{\commentMaxim}[1]{{\color{red}{\it ~Maxim:~}\tt #1}}	%comments
\renewcommand{\thefootnote}{\roman{footnote}}

\title{Topological states in arrays of optical waveguides engineered via mode interference}

\author{Roman~S.~Savelev}
\affiliation{ITMO University, Saint Petersburg 197101, Russia}

\author{Maxim~A.~Gorlach}
\email{m.gorlach@metalab.ifmo.ru}
\affiliation{ITMO University, Saint Petersburg 197101, Russia}

\begin{abstract}
Photonic structures with topologically nontrivial bands are usually designed by arranging simple meta-atoms, ideally, single-mode ones, in a carefully designed photonic lattice with symmetry that guarantees the emergence of topological states. Here we investigate an alternative option that does not require complex lattice geometry but instead relies on the  tuning of the parameters of the individual meta-atoms to achieve the degeneracy of the modes with different symmetry. As an illustrative example, we consider a one-dimensional array of equidistant identical periodic nanophotonic waveguides supporting degenerate modes with strongly asymmetric near field profiles giving rise to the coupling modulation. Exploiting this feature, we demonstrate that the proposed system supports topological edge modes and can be viewed as a generalization of the paradigmatic Su-Schrieffer-Heeger model, reducing to it for the suitable parameter choice. Our results thus provide an avenue to engineer topological states via mode interference which further expands the plethora of topological structures available in photonics being especially promising for nonlinear topological systems.
\end{abstract}

\maketitle

{\it Introduction}~--~Photonic topological structures feature a variety of striking physical properties, one-way disorder-robust propagation being the primary example~\cite{Lu2014,Lu2016,Khanikaev17,Ozawa_RMP}. Inherent robustness of electromagnetic topological modes to defects and imperfections holds further promises for topologically protected reflectionless waveguide bends, circulators and other on-chip photonic topological circuitry~\cite{Blanco-Redondo2020}.

To achieve such exciting functionalities at optical wavelengths, it has been proposed to utilize the arrays of laser-written optical waveguides~\cite{Szameit-2010}. In many cases, when the interaction of the nearest neighbors plays the dominant role, such arrays are well-described by the electromagnetic analogue of tight-binding model~\cite{Yariv}, where the coupling constant is determined by the overlap of the evanescent tails of waveguide modes. Furthermore, in the paraxial regime, waveguide lattices serve as simulators of quantum physics since the coordinate along the waveguide axis corresponds to time variable in Schr{\"o}dinger equation, the propagation constant plays the role of energy, while the refractive index modulation is analogous to the external potential~\cite{Rechtsman}. 

Such waveguide systems enabled straightforward realization of the variety of topological systems including one-dimensional Su-Schrieffer-Heeger model~\cite{Blanco-PRL,Blanco-Science} [Fig.~\ref{fig:Array}(a-c)], ``photonic graphene''~\cite{Plotnik}, photonic analogue of quantum Hall effect~\cite{Rechtsman}, optical realization of Weyl points~\cite{Noh-17} and higher-order topological phases~\cite{Noh-18}.

Quite interestingly, in all these situations the waveguides are considered as single-mode ones, and the topological properties of the bands are largely governed by the chosen lattice geometry. At the same time, the topological physics of the arrays of interacting multimode waveguides or meta-atoms remains vastly unexplored with only few first studies available currently~\cite{Bobylev2019,Vicencio2020}. 

In this Rapid Communication, we exploit this degree of freedom, studying the waveguides with two nearly degenerate modes in the frequency range of interest. Taking linear combinations of these modes, we recover strongly asymmetric near field distributions which give rise to the coupling modulation in the array as illustrated in Fig.~\ref{fig:Array}(d-f). We find out that as a consequence of this, the lattice of such two-mode waveguides can support topological states.

As a simple but illuminating example, we analyze an equidistant array of two-mode waveguides and demonstrate the existence of topological edge states at {\it both} edges of the array simultaneously. Furthermore, as we prove, the physics of the array is governed by the well-celebrated  Su-Scherieffer-Heeger model (SSH)~\cite{Su} under the suitable parameter choice. Note that previously SSH model has been realized in waveguide arrays by creating a lattice with two alternating distances between the neighboring waveguides~\cite{Malkova:09,Zhu2015,Blanco-PRL,Bleckmann}.

While in the linear case our approach provides just an alternative to the commonly accepted one based on lattice engineering, its advantages become more evident in the nonlinear regime. Indeed, on-site nonlinearities are relatively easy to control even at optical wavelengths, whereas the control over the coupling constants is highly challenging. Currently, tunable nonlinear couplings are only achieved at microwave frequencies via nonlinear varactor diode insertions~\cite{Hadad-Nature,Dobrykh,Serra-Garcia-Huber,Fleury}. Therefore, we expect our proposal not only to enrich the variety of photonic topological structures operating in the linear regime but also to bring  conceptual advances in the design of nonlinear topological systems operating at optical frequencies~\cite{Smirnova-Kruk}.

\begin{widetext}
%%%%%%%%%%%%%%%%%%%%%%%%%%%%%%%%%
   \begin{center}
    \begin{figure}[ht]
    \includegraphics[width=1.0\linewidth]{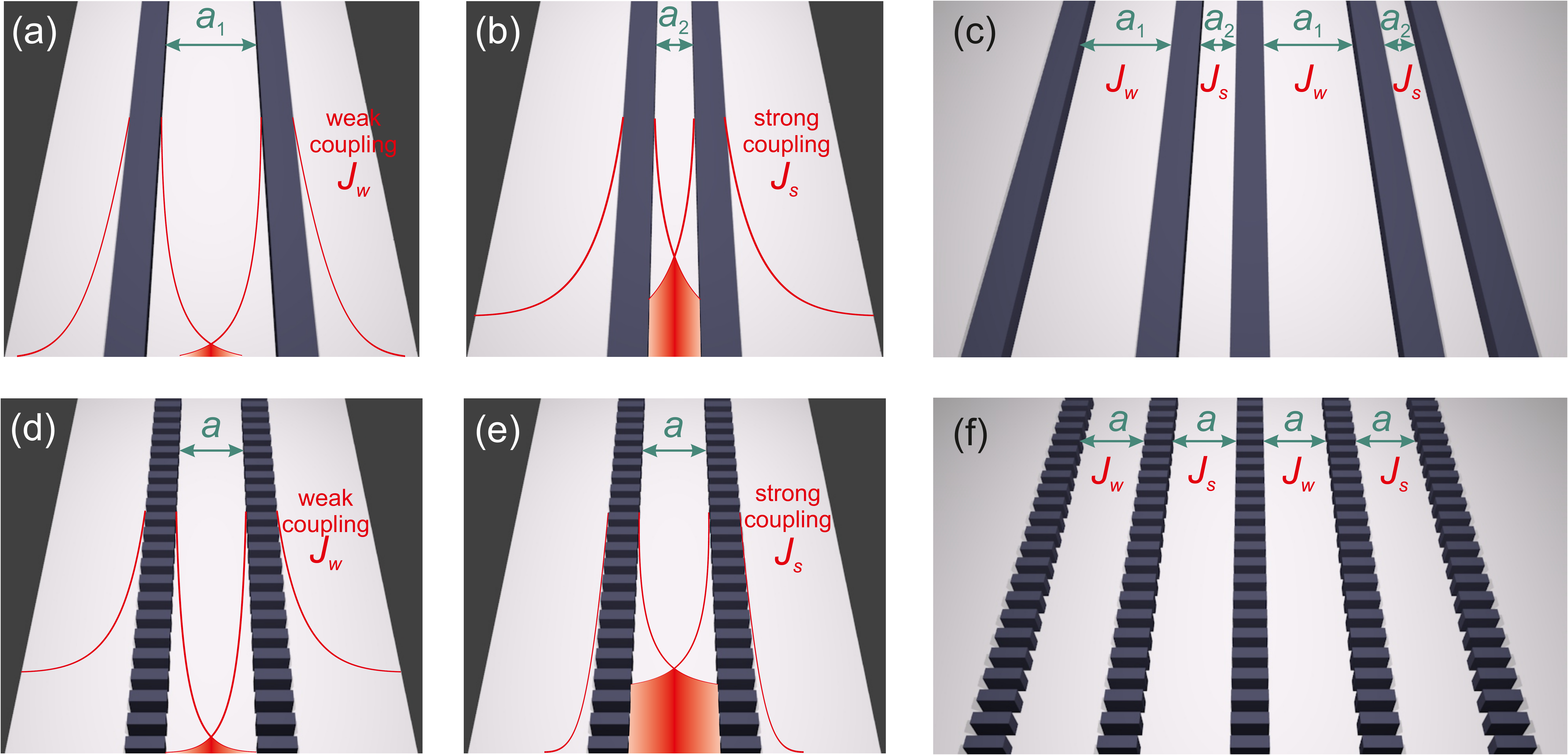}
    \caption{(a-c) Conventional design of the Su-Schrieffer-Heeger (SSH) model based on single-mode waveguides. Weak and strong coupling is achieved by increasing (a) or decreasing (b) the distance between the waveguides, respectively. The unit cell of the lattice (c) contains two waveguides. (d-f) Proposed realization of SSH model based on two-mode waveguides. Asymmetry of the field distribution allows to tailor the couplings without varying the distance between the waveguides. The resultant lattice (f) contains a single waveguide in the unit cell.} 
    \label{fig:Array}
    \end{figure}
    \end{center}
%%%%%%%%%%%%%%%%%%%%%%%%%%%%%%%%%
\end{widetext}

{\it Theoretical model}~--~An array of waveguides with two degenerate modes Fig.~\ref{fig:Array}(f) can be described by the following set of coupled mode equations (Supplementary Materials, Sec.~I):
\begin{equation}\label{CoupledModeGeneral}
-i\,\frac{d}{dz}\,
\begin{pmatrix}
u_n\\
v_n
\end{pmatrix}
=\hat{\kap}_{n,n-1}\,
\begin{pmatrix}
u_{n-1}\\
v_{n-1}
\end{pmatrix}+
\hat{\kap}_{n,n+1}\,
\begin{pmatrix}
u_{n+1}\\
v_{n+1}
\end{pmatrix}\:,
\end{equation}
where $z$ axis corresponds to the propagation direction, $u_n$ and $v_n$ denote the amplitudes of the modes supported by $n^{\rm{th}}$ waveguide which are assumed to be weakly dependent on $z$, and $\hat{\kap}_{n,n\pm 1}$ are the matrices describing the coupling between the modes of the neighboring waveguides. Based on the symmetry of mode profiles, coupling matrices have the following general structure (see Supplementary Materials, Sec.~III for details):
\begin{equation}\label{CouplingMatrixFull}
\hat{\kap}_{n,n\pm 1}=
\begin{pmatrix}
\kap & \pm i\,\Delta\\
\pm i\,\Delta & -\gamma
\end{pmatrix}\:,
\end{equation}
where the equality $\kap_{n,n\pm 1}^{(12)}=\kap_{n,n\pm 1}^{(21)}$ is ensured by the appropriate relative normalization of the modes $u_n$ and $v_n$.

In the case of a periodic array of waveguides, Bloch theorem yields that $\left(u_n, v_n\right)=e^{ik n}\,\left(u, v\right)$, where $k$ is the Bloch wave number in the direction of periodicity $x$, while $z$ dependence of the amplitudes is captured by the factor $e^{i\delta k_z z}$. Here, $\delta k_z$ is the difference between the propagation constants of the array eigenmode and the eigenmode of an isolated waveguide. With these assumptions, we derive an eigenvalue equation of the form:
\begin{equation}\label{CoupledModeEig}
\delta k_z\,
\begin{pmatrix}
u\\
v
\end{pmatrix}
=\hat{H}(k)\,
\begin{pmatrix}
u\\
v
\end{pmatrix}\:,
\end{equation}
where Bloch Hamiltonian $\hat{H}(k)$ is given by
\begin{equation}\label{BlochHamiltonian}
\hat{H}(k)=
\begin{pmatrix}
2\kap\,\cos k & -2\Delta\, \sin k\\
-2\Delta\, \sin k & -2\gamma\,\cos k
\end{pmatrix}
\:,
\end{equation}
and the associated spectrum reads:
\begin{equation}\label{Spectrum}
\delta k_z^{(\pm)}=\left(\kap-\gamma\right)\,\cos k\pm \sqrt{(\kap+\gamma)^2\,\cos^2 k+4\Delta^2\,\sin^2 k}\:,
\end{equation}
where Bloch wave number $k$ ranges from $-\pi$ to $\pi$. For $\kap$ and $\gamma$ of different signs the bandgap common for all wave numbers $k$ does not exist either because of $\delta k_z^{(+)}(0)<\delta k_z^{(-)}(\pi)$ or $\delta k_z^{(+)}(\pi)<\delta k_z^{(-)}(0)$. 

To investigate the topological properties of our system for $\kap$ and $\gamma$ of the same sign, we first consider the special case $\kap=\gamma$. In this situation, the spectrum Eq.~\eqref{Spectrum} appears to be symmetric with respect to zero energy and Bloch Hamiltonian Eq.~\eqref{BlochHamiltonian} possesses chiral symmetry. As a result of that, the Hamiltonian can be converted to the off-diagonal form via unitary transformation $\hat{P}=(\hat{I}+i\,\sigma_x)/\sqrt{2}$:
\begin{equation}\label{OffDiagonal}
\begin{split}
& \hat{H}'(k)\equiv \hat{P}\,\hat{H}(k)\,\hat{P}^{-1}\\
=
& 
\begin{pmatrix}
0 & -2i\kap\,\cos k-2\Delta\,\sin k\\
2i\kap\,\cos k-2\Delta\,\sin k & 0
\end{pmatrix}
\:.
\end{split}
\end{equation}
Straightforward evaluation of winding number yields $\mathcal{W}=1$ indicating topological nature of energy bands.

In the general case $\kap\not=\gamma$, the argument above is no longer valid and chiral symmetry is seemingly broken. To overcome this difficulty and define the topological invariant, we consider an {\it extended unit cell} including two waveguides. Since each of the waveguides supports two modes, Bloch Hamiltonian becomes $4\times 4$: 
\begin{equation}
\hat{H}_{\rm{ext}}(k)=\begin{pmatrix}
0 & \hat{h}(k)\\
\hat{h}^\dag(k) & 0
\end{pmatrix}
\:,
\end{equation}
where single $2\times 2$ block is given by
\begin{equation}
\hat{h}=\begin{pmatrix}
\kap\,(e^{-ik}+1) & -i\Delta\,(e^{-ik}-1)\\
-i\Delta\,(e^{-ik}-1) & -\gamma\,(e^{-ik}+1) 
\end{pmatrix}
\end{equation}
and $k$ varies again from $-\pi$ to $\pi$. As a result of period doubling, two original dispersion curves fold into four bands. In folded bands representation, chiral symmetry of Bloch Hamiltonian is recovered. This allows us to evaluate the winding number using the standard procedure~\cite{Ryu}, which yields that the system is topological with $\mathcal{W}=1$ for any $\kap$ and $\gamma$ of the same sign, provided $\Delta\not=0$. Note that the similar conclusion can be obtained by the direct calculation of the Zak phase, without using an extended unit cell (Supplementary Materials, Sec.~IV).

To provide further insights into the topological properties of the proposed system, we demonstrate now that in the limit $\kap=\gamma$ the physics of equidistant array of two-mode waveguides is captured by the Su-Schrieffer-Heeger model. To this end, we perform a unitary transformation with the matrix
\begin{equation}
\hat{U}=\frac{1}{\sqrt{2}}\,
\begin{pmatrix}
1 & i & 0 & 0\\
0 & 0 & 1 & -i\\
1 & -i & 0 & 0\\
0 & 0 & 1 & i
\end{pmatrix}
\:,
\end{equation} 
which brings the Hamiltonian to the block-diagonal form
\begin{equation}
\hat{H}_{\rm{block}}=\begin{pmatrix}
\hat{H}_{+}(k) & 0\\
0 & \hat{H}_{-}(k)
\end{pmatrix}\:,
\end{equation}
where each of the $2\times 2$ blocks is given by
\begin{small}
\begin{equation}
\hat{H}_{\pm}=
\begin{pmatrix}
0 & \mp\Delta+\kap+(\pm\Delta+\kap)\,e^{-ik}\\
\mp\Delta+\kap+(\pm\Delta+\kap)\,e^{ik} & 0
\end{pmatrix}
\:.
\end{equation}
\end{small}
Such structure of the blocks corresponds to the Bloch Hamiltonian of the Su-Schrieffer-Heeger model (SSH) with coupling constants $J_1=\kap-\Delta$ and $J_2=\kap+\Delta$, $J_1$ being an intracell coupling for $\hat{H}_{+}$ block and inter-cell coupling for $\hat{H}_{-}$ block. Hence, our system of two-mode waveguides with $\kap=\gamma$ can be presented as two uncoupled SSH arrays with the  different dimerizations. Since SSH model supports an edge state only at the weak link edge, our system should support localized states at both edges simultaneously  regardless of the number of waveguides comprising the array.

%\begin{widetext}
%%%%%%%%%%%%%%%%%%%%%%%%%%%%%%%%%
   \begin{center}
    \begin{figure*}[ht]
    \includegraphics[width=1.0\linewidth]{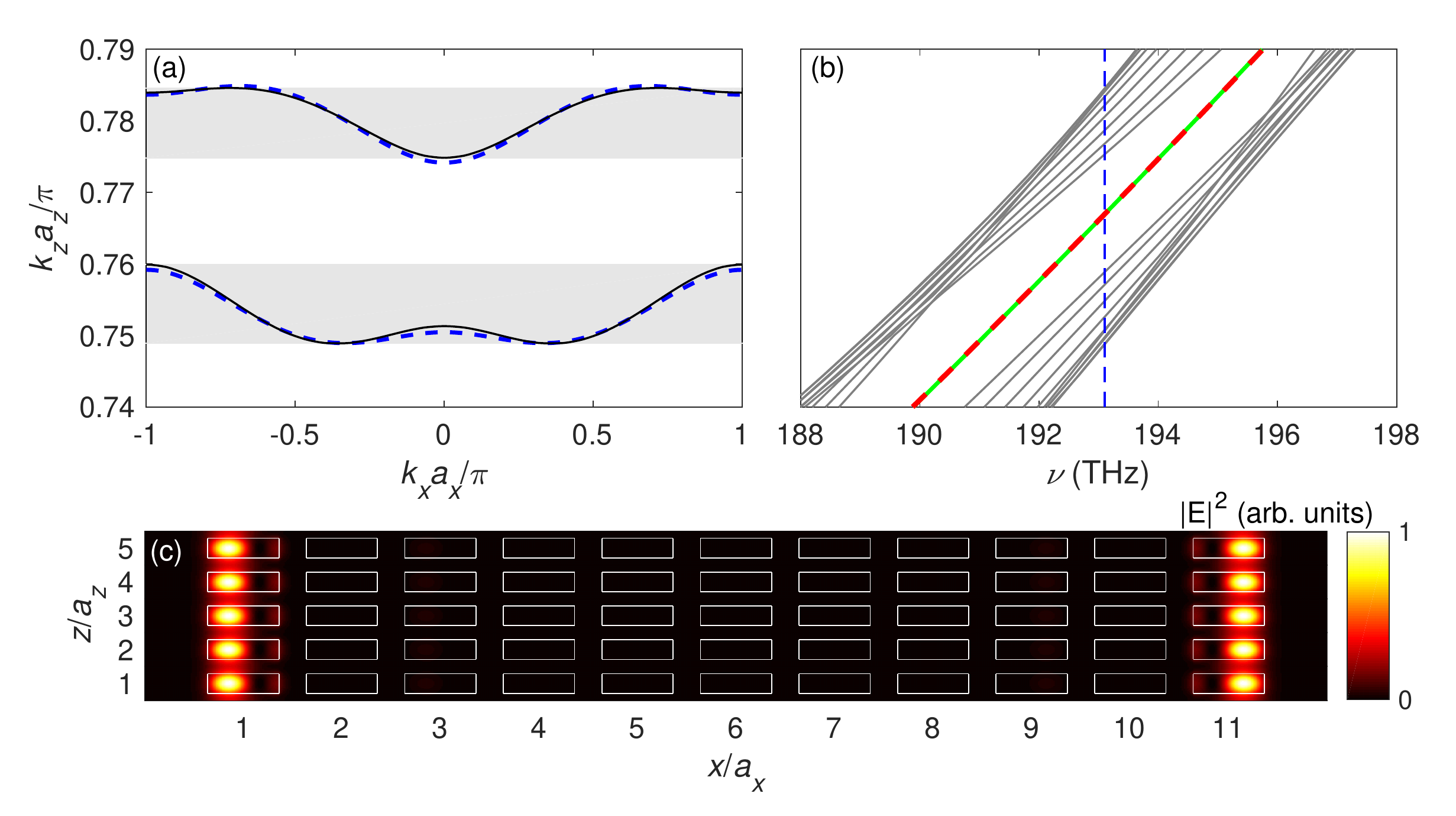}
    \caption{(a) Dispersion of the modes $k_z(k_x)$ in the infinite array of periodic waveguides with the periods $a_x=830$~nm and $a_z=285$~nm along $x$ and $z$ directions, respectively. The constituent nanoparticles have the shape of rectangular parallelepiped  with the dimensions $d_x=600$~nm, $d_y=660$~nm and $d_z=165$~nm. The results are calculated for a fixed frequency $\nu=193.1$~THz using CST Microwave Studio software package. Shaded grey areas show the allowed bands. (b) Propagation constants of the modes supported by the array of 11 waveguides periodic in $z$ direction versus excitation frequency $\nu$. Bulk modes are shown by solid grey lines, edge modes~-- by dashed red and solid green curves. Vertical dashed line indicates frequency chosen for the calculations in panels (a) and (c). (c) Electric field intensity for the edge mode. Five unit cells along $z$ direction are shown. White rectangles indicate the constituent nanoparticles of periodic waveguides. Both edge modes feature almost identical intensity distributions but differ in phase (see Supplementary Materials, Sec.~V).} 
    \label{fig:Numerics}
    \end{figure*}
    \end{center}
%%%%%%%%%%%%%%%%%%%%%%%%%%%%%%%%%
%\end{widetext}

{\it Numerical simulations}~--~To confirm our theoretical predictions, we propose a realistic design of waveguide array operating at telecom frequencies. We perform full-wave numerical simulations of the proposed system [Fig.~\ref{fig:Array}(f)] assuming that the periodic waveguides with subwavelength period are made of crystalline silicon with refractive index $n=3.5$ deposited on the glass substrate with $n=1.45$. Other parameters of the system are chosen in such a way that the degeneracy of HE$_{11}$ and TE$_{01}$ modes of a single waveguide is achieved at frequency $\nu\approx193.1$~THz, which corresponds to vacuum wavelength around $1.55\,\mu\rm{m}$. Note also that the two waveguide modes HE$_{11}$ and TE$_{01}$ have different behavior with respect to reflection in $yz$ plane~\cite{Savelev-JAP} which is necessary to realize the pattern of alternating couplings via  constructive and destructive mode interference.
\par

First, we examine the modes of an infinite array of identical  waveguides at fixed frequency. Since in the propagating geometry $k_z$ plays the role analogous to energy in the Schr{\"o}dinger equation, it is instructive to plot the dependence $k_z(k_x)$, where $k_x$ is  Bloch wave number characterizing phase advance between the neighboring waveguides (note that $k$ in the theoretical model corresponds to $k_x a_x$, where $a_x$ is the distance between the adjacent waveguides). Figure~\ref{fig:Numerics}(a) suggests that the interaction between the adjacent waveguides gives rise to the two distinct bands separated by the complete bandgap. Note, however, that the calculated dispersion is different from that expected for the SSH model, because $\kap\not=\gamma$ in this design. More precisely, $\kap=0.0265/a_z$, $\gamma=0.0118/a_z$, $\Delta=0.0263/a_z$, where $a_z$ is the period of the waveguide along $z$ axis, i.e. $\kap$ and $\gamma$ coupling constants differ approximately by 2 times.

 %\red{$\kap/k_{z0}=0.011$, $\gamma/k_{z0}=0.0049$, $\Delta/k_{z0}=0.0109$, where $k_{z0}$ is the Bloch wavenumber of an isolated waveguide at the frequency $\nu=193.1$~THz. I.e. $\kap$ and $\gamma$ coupling constants differ approximately by 2 times.}

% $\kap=0.0239$, $\gamma=0.00762$, $\Delta=0.0236$,
% Second, the waveguides are spaced quite closely with half-wavelength  distance between them. Therefore, next-nearest neighbor coupling may also have effect on the dispersion. \red{Should we write here something about the influence of the lower TM band, mentioning that we are dealing with 3-band system?}

Switching to the finite array of 11 waveguides, we observe two modes with the propagation constants $k_z$ within the bandgap of the periodic structure which can be identified as edge-localized states. Interestingly, these in-gap states are quite stable against the change of excitation frequency [Fig.~\ref{fig:Numerics}(b)] perfectly overlapping with each other in the range from 190 to 196~THz and existing inside the bandgap even in a broader range $185-200$~THz.

Examining the field distribution of the edge-localized modes, we observe that they localize at {\it both} edges of the array since there is no physical distinction between the two edges. This conclusion perfectly agrees with our theoretical model, since the waveguide array can be viewed as two uncoupled SSH models with the different dimerizations hosting the localized states on the opposite edges.

Due to symmetry, edge eigenmodes are either symmetric or antisymmetric with respect to reflection relative to the array center and have similar field distributions depicted in Fig.~\ref{fig:Numerics}(c). As further elaborated in Supplementary Materials, Sec.~V, the splitting between these modes $\delta\nu = \Delta\nu/\nu$ depends on frequency and parity of the number $N$ of the waveguides and for 11 waveguides it is of the order of $10^{-5}$. Since the splitting is caused by the overlap of the two edge-localized evanescent field tails, it decays exponentially with $N$. This means that if the mode is locally excited at one of the edges, it will remain localized at the same edge at propagation distances up to $z\approx \frac{v_g}{2\pi\,\Delta\nu}$, where $v_g$ is a group velocity of the waveguide mode, and this distance can be made large enough by increasing $N$.

Another important feature of the considered system is that the wave propagating along either edge is spin-polarized in the sense that its field distribution is characterized by the non-zero transverse $y$ component of the total spin. Although even and odd edge modes have necessarily zero total $y$ component of the spin, their linear combinations localized at a single edge do exhibit a preferential direction of polarization rotation. To capture this property, we calculate the total spin of the superposition of two eigenmodes $a_n=\dfrac{1}{\sqrt{2}}(u_n + v_n)$ using the following formula~\cite{Bliokh}:
% \begin{smallmatrix}\text{half of the}\\\text{unit cell}\end{smallmatrix
\begin{equation}
\VB{S} = \int\limits_{\text{unit cell}} \dfrac{1}{4}\,\mathrm{Im}\left[\eps_0\eps(\VB{r})\VB{E}\times\VB{E}^* + \mu_0\VB{H}\times\VB{H}^* \right] \,dV,
\end{equation}
where electric and magnetic fields ${\bf E}$ and ${\bf H}$ of the mode are normalized in such a way, that the total energy of the mode $a_n$ in a single unit cell is equal to $\hbar\,\omega$. The obtained value $S_x\approx0.13$ indicates that fields at the left (right) edge exhibit on average elliptical polarization and rotate preferentially counterclockwise (clockwise). Thus, they can asymmetrically couple to the circularly polarized dipole sources placed near the edge waveguide.

{\it Conclusions}~--~In conclusion, we have demonstrated that topological states can be tailored not only through the lattice design, but also by engineering the modes of a single waveguide. As we have proved, exploiting the interference of the two sets of modes, one can realize topological states even in the simplest lattice geometries, and our simulations show the feasibility of this route.

We envision that the developed approach will be beneficial for tunable and nonlinear topological structures, since it enables tuning of topological states by modifying on-site properties. Thus, our proposal enriches the toolkit to design the  topological structures opening further avenues to tune and reconfigure them.

{\it Acknowledgments}~--~We acknowledge valuable discussions with Alexander Poddubny and Daria Smirnova. This work was supported by the Russian Science Foundation: theoretical models were developed with the support by the Grant No.~20-72-10065, numerical simulations were supported by the Grant No.~19-72-10129. M.A.G. acknowledges partial support by the Foundation for the Advancement of Theoretical Physics and Mathematics ``Basis''.

\bibliography{TopologicalLib}

\end{document}

\newcommand{\e}{{\rm e}}
\newcommand{\rmi}{{\rm i}}
\renewcommand{\Im}{\mathop\mathrm{Im}\nolimits}
\newcommand{\red}[1]{\textcolor{red}{#1}}
\newcommand{\blue}[1]{{\color{blue}#1}}
\newcommand{\rot}[1]{\text{rot}\, #1}
\newcommand{\bas}[1]{\hat{{\bf #1}}}
\newcommand{\refPR}[1]{[\onlinecite{#1}]}
\newcommand{\cra}[1]{\hat{a}^{\dag}_{#1}}  %creation operator a
\newcommand{\ana}[1]{\hat{a}_{#1}^{\vphantom{\dag}}}         %annihilation operator a
\newcommand{\num}[1]{\hat{n}_{#1}}         %annihilation operator a
\newcommand{\bra}[1]{\left|#1\right>}      %Dirac designations
\newcommand{\ket}[1]{\left<#1\right|}
\newcommand{\eps}{\varepsilon}      %Greek epsilon
\newcommand{\om}{\omega}      %Greek omega
\newcommand{\kap}{\varkappa}      %Greek kappa

\newcommand{\spm}[1]{#1^{(\pm)}}
\newcommand{\skvk}[2]{\left<#1\left|\frac{\partial #2}{\partial k}\right.\right>} %scalar products
\newcommand{\skvv}[2]{\left<#1\left|#2\right.\right>}
\newcommand{\df}[2]{\frac{\partial #1}{\partial #2}}
\newcommand{\ds}[1]{\partial #1/\partial k}

%%%%%%%%%%%%%%%%%%%%%%%%%%%%%%%%%%%%%%%%%%%%

\renewcommand{\thefigure}{S\arabic{figure}}
\renewcommand{\theequation}{S\arabic{equation}}
\title{Topological states in arrays of optical waveguides engineered via mode interference. Supplementary Materials}
\author{Roman~S.~Savelev}
\affiliation{ITMO University, Saint Petersburg 197101, Russia}
\author{Maxim~A.~Gorlach}
\email{m.gorlach@metalab.ifmo.ru}
\affiliation{ITMO University, Saint Petersburg 197101, Russia}

\maketitle
\tableofcontents

%%%%%%%%%%%%%%%%%%%%%%%%%%%%%%%%%%%%%%%%%%%%
\section{Coupled mode theory for two-mode waveguides}\label{sec:AppA}

We consider an array that consists of waveguides supporting two nearly degenerate modes in the frequency range of interest. The modes of $n^{\rm{th}}$ waveguide with permittivity distribution $\eps_n({\bf r})$ satisfy Maxwell's equations
\begin{align}
& \rot{{\bf E}_n^{(1)}}=i\,q\,{\bf H}_n^{(1)}\:, & \rot{{\bf E}_n^{(2)}}=i\,q\,{\bf H}_n^{(2)}\:,\label{Wav1}\\
& \rot{{\bf H}_n^{(1)}}=-i\,q\,\eps_n({\bf r})\,{\bf E}_n^{(1)}\:, & \rot{{\bf H}_n^{(2)}}=-i\,q\,\eps_n({\bf r})\,{\bf E}_n^{(2)}\:,\label{Wav2}
\end{align}
where CGS system of units and $e^{-i\om\,t}$ time convention are used, $q=\om/c$ and superscript indices 1 and 2 label the two modes of the same waveguide. The modes of waveguide array in turn are found from the equations
\begin{equation}\label{Array}
\rot{{\bf E}}=i\,q\,{\bf H}\:,\mspace{8mu} \rot{{\bf H}}=-i\,q\,\eps({\bf r})\,{\bf E}\:,
\end{equation}
where $\eps({\bf r})$ is the permittivity distribution in the entire array consisting of waveguides arranged in  uniform background medium with permittivity $\eps_b$, i.e. $\eps({\bf r})=\eps_b+\sum_n\,\left[\eps_n({\bf r})-\eps_b\right]$. We search the modes of waveguide array in the form
\begin{equation}\label{Ansatz}
\begin{pmatrix}
{\bf E}\\
{\bf H}
\end{pmatrix}
=\sum\limits_n\,a_n^{(1)}(z)\,
\begin{pmatrix}
{\bf E}_n^{(1)}\\
{\bf H}_n^{(1)}
\end{pmatrix}+
\sum\limits_n\,a_n^{(2)}(z)\,
\begin{pmatrix}
{\bf E}_n^{(2)}\\
{\bf H}_n^{(2)}
\end{pmatrix}\:,
\end{equation}
where $a_n^{(1,2)}(z)$ are the amplitudes of the modes in coupled waveguides which depend on $z$ coordinate along the waveguide axis. Combining Eqs.~\eqref{Array}, \eqref{Ansatz} and taking into account Eqs.~\eqref{Wav1}, \eqref{Wav2}, we derive the following relations for the amplitudes $a_n^{(1,2)}(z)$:
\begin{gather}
\sum\limits_n\,\left\lbrace \frac{da_n^{(1)}}{dz}\,\left[\bas{z}\times{\bf E}_n^{(1)}\right]+\frac{da_n^{(2)}}{dz}\,\left[\bas{z}\times{\bf E}_n^{(2)}\right]\right\rbrace=0\:,\label{A1E0}\\
\sum\limits_n\,\left\lbrace \frac{da_n^{(1)}}{dz}\,\left[\bas{z}\times{\bf H}_n^{(1)}\right]+\frac{da_n^{(2)}}{dz}\,\left[\bas{z}\times{\bf H}_n^{(2)}\right]\right\rbrace=-iq\,\sum\limits_n\,\left[\eps({\bf r})-\eps_n({\bf r})\right]\,\left[a_n^{(1)}\,{\bf E}_n^{(1)}+a_n^{(2)}\,{\bf E}_n^{(2)}\right]\:,\label{A2E0}
\end{gather}
where $\bas{z}$ is a unit vector along the waveguide axis. Next we take  scalar products of  Eq.~\eqref{A1E0} by ${\bf H}_m^{(1,2)*}$ and Eq.~\eqref{A2E0} by ${\bf E}_m^{(1,2)*}$. Substracting these two equations from each other and integrating over the volume of the unit cell, we get
\begin{gather}
\sum\limits_n\,\frac{da_n^{(1)}}{dz}\,\bas{z}\cdot\int\,\left(\left[{\bf E}_n^{(1)}\times{\bf H}_m^{(1)*}\right]+\left[{\bf E}_m^{(1)*}\times{\bf H}_n^{(1)}\right]\right)\,dV\notag\\
+\sum\limits_n\,\frac{da_n^{(2)}}{dz}\,\bas{z}\cdot\int\,\left(\left[{\bf E}_n^{(2)}\times{\bf H}_m^{(1)*}\right]+\left[{\bf E}_m^{(1)*}\times{\bf H}_n^{(2)}\right]\right)\,dV\notag\\
=iq\,\sum\limits_n\,a_n^{(1)}\,\int\,\left[\eps({\bf r})-\eps_n({\bf r})\right]\,{\bf E}_m^{(1)*}\cdot{\bf E}_n^{(1)}\,dV+iq\,\sum\limits_n\,a_n^{(2)}\,\int\,\left[\eps({\bf r})-\eps_n({\bf r})\right]\,{\bf E}_m^{(1)*}\cdot{\bf E}_n^{(2)}\,dV\label{A1Eq}\\
\sum\limits_n\,\frac{da_n^{(1)}}{dz}\,\bas{z}\cdot\int\,\left(\left[{\bf E}_n^{(1)}\times{\bf H}_m^{(2)*}\right]+\left[{\bf E}_m^{(2)*}\times{\bf H}_n^{(1)}\right]\right)\,dV\notag\\
+\sum\limits_n\,\frac{da_n^{(2)}}{dz}\,\bas{z}\cdot\int\,\left(\left[{\bf E}_n^{(2)}\times{\bf H}_m^{(2)*}\right]+\left[{\bf E}_m^{(2)*}\times{\bf H}_n^{(2)}\right]\right)\,dV\notag\\
=iq\,\sum\limits_n\,a_n^{(1)}\,\int\,\left[\eps({\bf r})-\eps_n({\bf r})\right]\,{\bf E}_m^{(2)*}\cdot{\bf E}_n^{(1)}\,dV+iq\,\sum\limits_n\,a_n^{(2)}\,\int\,\left[\eps({\bf r})-\eps_n({\bf r})\right]\,{\bf E}_m^{(2)*}\cdot{\bf E}_n^{(2)}\,dV\:,\label{A2Eq}
\end{gather}
where $m$ is an arbitrary integer. Note that the terms with $da_m^{(2)}/dz$ in Eq.~\eqref{A1Eq} and with $da_m^{(1)}/dz$ in Eq.~\eqref{A2Eq} vanish due to the  ortogonality of the modes of a single waveguide (see Sec.~\ref{sec:AppB} for details). Since the fields of waveguide modes decay exponentially outside of the waveguide, Eqs.~\eqref{A1Eq}, \eqref{A2Eq} can be further truncated to yield
\begin{equation}\label{CoupledMode}
\frac{d}{dz}\,
\begin{pmatrix}
a_m^{(1)}\\
a_m^{(2)}
\end{pmatrix}
=i\,\sum\limits_{l=\pm 1}
\begin{pmatrix}
\kap^{(11)}_{m,m+l} & \kap^{(12)}_{m,m+l}\\
\kap^{(21)}_{m,m+l} & \kap^{(22)}_{m,m+l}
\end{pmatrix}
\,
\begin{pmatrix}
a_{m+l}^{(1)}\\
a_{m+l}^{(2)}
\end{pmatrix}\:.
\end{equation}
The elements of the coupling matrix are defined as
\begin{equation}\label{CouplingMatrix}
\kap_{n,n+l}^{(pq)}=\frac{q}{\mathcal{P}^{(p)}}\,\int\,\left[\eps({\bf r})-\eps_{n+l}({\bf r})\right]\,{\bf E}_n^{(p)*}\cdot{\bf E}_{n+l}^{(q)}\,dV\:,
\end{equation}
where $\mathcal{P}_n^{(p)}=\bas{z}\cdot\int\,\left[{\bf E}_n^{(p)*}\times{\bf H}_n^{(p)}+{\bf E}_n^{(p)}\times{\bf H}_n^{(p)*}\right]\,dV$ is proportional to the power carried by the respective waveguide mode and can be set to unity by the proper normalization of the mode field; $p,q$ indices are equal to 1 or 2, $l=\pm 1$. As a consequence of Eq.~\eqref{CouplingMatrix}, the coupling strength $\kap_{n,n\pm 1}$ is mostly determined by the overlap of the fields of the two modes inside $n^{\rm{th}}$ waveguide.

Note also that the magnitude of the off-diagonal terms in the coupling matrix is affected by the relative normalization of the modes. Assume that $\kap_{nm}^{(pq)}=\kap_1$ and $\kap_{nm}^{(qp)}=\kap_2$. We redefine the field of the mode as $E^{(p)}=c\,\tilde{E}^{(p)}$. The coupling constants will also be redefined as follows:
\begin{equation}
\begin{split}
\tilde{\kap}^{(pq)}=\kap_1/c\:,\\
\tilde{\kap}^{(qp)}=\kap_2\,c\:.
\end{split}
\end{equation}
Hence, by choosing $c=\pm\sqrt{\kap_1/\kap_2}$, we can ensure that the coupling matrix for each pair of neighboring waveguides is symmetric, which is used in the main text.

\section{Orthogonality relation for two modes of a waveguide}\label{sec:AppB}

For the sake of completeness, in this Section we provide a proof that the two modes $\left({\bf E}^{(1)},{\bf H}^{(1)}\right)$ and $\left({\bf E}^{(2)},{\bf H}^{(2)}\right)$ of a given lossless waveguide having different propagation constants satisfy the following orthogonality relation~\cite{Okamoto}:
\begin{equation}\label{Orthogonal}
\bas{z}\cdot\int\,\left[{\bf E}^{(1)*}\times{\bf H}^{(2)}+{\bf E}^{(2)}\times{\bf H}^{(1)*}\right]\,dV=0\:,
\end{equation}
where the integration is carried on over the unit cell of the periodic waveguide. To show this, we start from the quantity
\begin{equation}
\begin{split}
I=& \int \nabla\cdot\left[{\bf E}^{(1)*}\times{\bf H}^{(2)}+{\bf E}^{(2)}\times{\bf H}^{(1)*}\right]dV=\int\left({\bf H}^{(2)}\cdot\rot{{\bf E}^{(1)*}}-{\bf E}^{(1)*}\cdot\rot{{\bf H}^{(2)}}+{\bf H}^{(1)*}\cdot\rot{{\bf E}^{(2)}}-{\bf E}^{(2)}\cdot\rot{{\bf H}^{(1)*}}\right)dV\\
&\overset{\text{Eq.}~\eqref{Wav1},\eqref{Wav2}}{=}iq\,\int\,\left(-{\bf H}^{(2)}\cdot{\bf H}^{(1)*}+\eps({\bf r})\,{\bf E}^{(1)*}\cdot{\bf E}^{(2)}+{\bf H}^{(1)*}\cdot{\bf H}^{(2)}-\eps^*({\bf r})\,{\bf E}^{(2)}\cdot{\bf E}^{(1)*}\right)dV=0\:,
\end{split}
\label{I=0}
\end{equation}
since $\eps({\bf r})$ is purely real in the lossless case. Next we present the fields of eigenmodes in the form
\begin{equation}
{\bf E}^{(1,2)}({\bf r})=\tilde{{\bf E}}^{(1,2)}({\bf r})\,e^{i\beta_{1,2}\,z}\:,\mspace{8mu}{\bf H}^{(1,2)}({\bf r})=\tilde{{\bf H}}^{(1,2)}({\bf r})\,e^{i\beta_{1,2}\,z}\:,
\end{equation}
which yields for the integral $I$
\begin{equation}\label{Ieval}
\begin{split}
&I=\int\,\nabla\cdot\left[\left(\tilde{{\bf E}}^{(1)*}\times\tilde{{\bf H}}^{(2)}+\tilde{{\bf E}}^{(2)}\times\tilde{{\bf H}}^{(1)*}\right)\,e^{i(\beta_2-\beta_1)\,z}\right]\,dV\\
&=e^{i(\beta_2-\beta_1)\,z}\,\left[\int\,\nabla\cdot\left(\tilde{{\bf E}}^{(1)*}\times\tilde{{\bf H}}^{(2)}+\tilde{{\bf E}}^{(2)}\times\tilde{{\bf H}}^{(1)*}\right)\,dV+
i\,(\beta_2-\beta_1)\,\bas{z}\cdot\int\,\left({\bf E}^{(1)*}\times{\bf H}^{(2)}+{\bf E}^{(2)}\times{\bf H}^{(1)*}\right)\,dV\right]=\\
&=e^{i(\beta_2-\beta_1)\,z}\,\int\,\left(\tilde{{\bf E}}^{(1)*}\times\tilde{{\bf H}}^{(2)}+\tilde{{\bf E}}^{(2)}\times\tilde{{\bf H}}^{(1)*}\right)\cdot{\bf n}\,dS+i\,(\beta_2-\beta_1)\,\bas{z}\cdot\int\,\left({\bf E}^{(1)*}\times{\bf H}^{(2)}+{\bf E}^{(2)}\times{\bf H}^{(1)*}\right)\,dV\:.
\end{split}
\end{equation}
The surface integral in the last expression vanishes due to the periodicity of the fields in $z$ direction and rapid decay in transverse directions. On the other hand, from~\eqref{I=0} $I=0$. Thus, we recover the identity Eq.~\eqref{Orthogonal}. Clearly, Eq.~\eqref{Orthogonal} is still valid for the two modes with very close values of propagation constants.

\section{Structure of the coupling matrices}

\begin{center}
    \begin{figure}[hb]
    \includegraphics[width=0.6\linewidth]{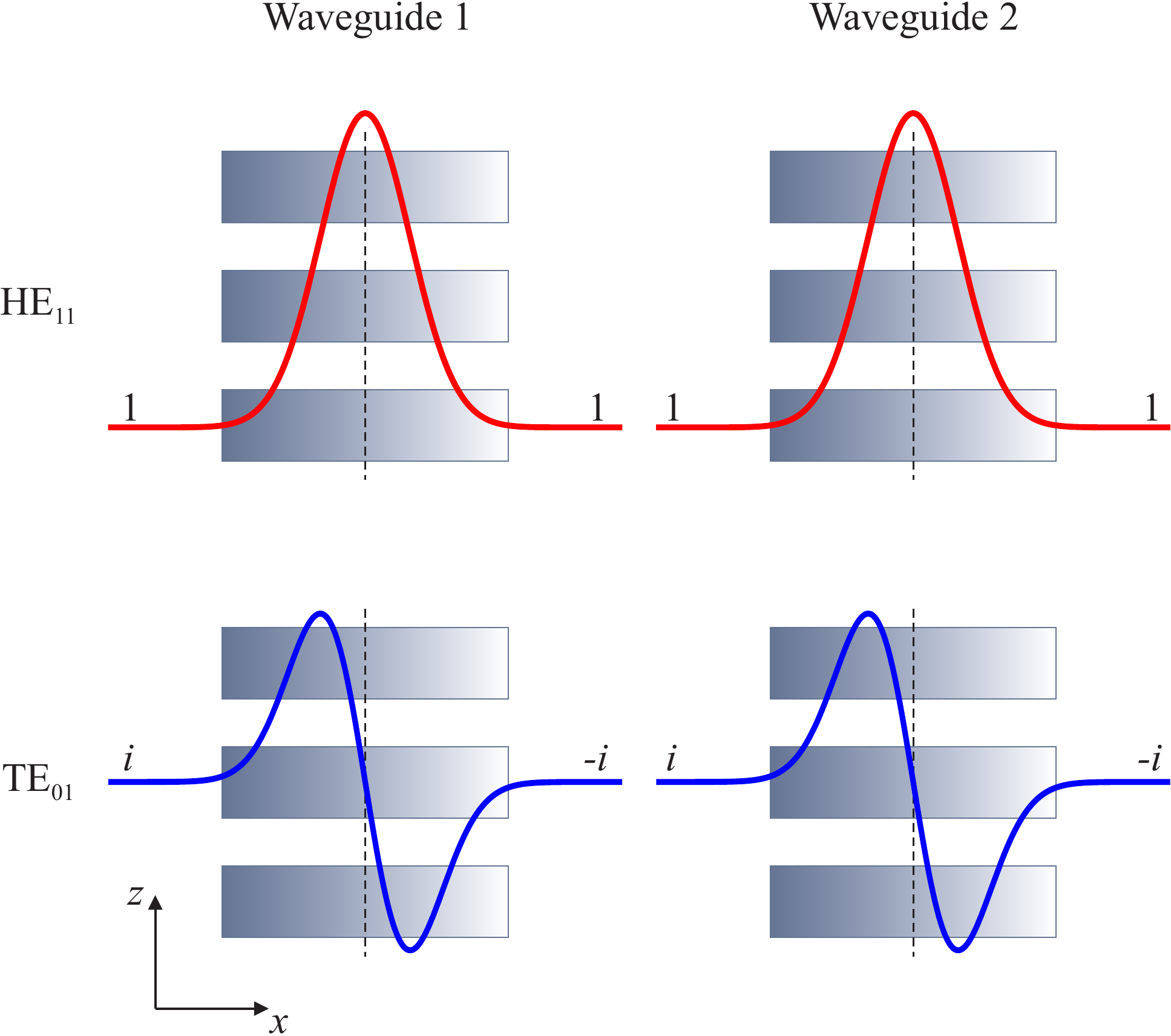}
    \caption{Mode coupling in the array of waveguides supporting a pair of degenerate modes. Curves illustrate the magnitude of electric field $E_y$ component, numbers near each curve indicate  the phase of the field for the respective mode.} 
    \label{fig:Coupling}
    \end{figure}
    \end{center}

In this Section, we argue that the coupling matrices for two-mode waveguides under study indeed have the structure Eq.~(2) of the article main text. We adopt a semi-qualitative approach assuming in accordance with Eq.~\eqref{CouplingMatrix} that the coupling is governed by the scalar product of electric fields of the modes inside the respective waveguide. As further detailed in Sec.~\ref{sec:AppE}, the dominant component of electric field is $E_y$ for both modes, where the spatial profile of $E_y$ is sketched in Fig.~\ref{fig:Coupling} by solid curves.

Therefore, taking into account that $\kap_{12}^{(pq)}\propto E_{1y}^{(p)*}\,E_{2y}^{(q)}$, we recover the following structure of the coupling matrix:
\begin{equation}\label{Kappa12}
    \hat{\kap}_{12}=
    \begin{pmatrix}
    \kap & i\Delta\\
    i\Delta & -\gamma
    \end{pmatrix}
    \:,
\end{equation}
where the precise values of real positive parameters $\kap$, $\gamma$ and $\Delta$ are either extracted from numerical simulations of the field distribution, or from fitting the numerical data on the dispersion of the array modes. At the same time, $\kap_{21}^{(pq)}\propto E_{2y}^{(p)*}\,E_{1y}^{(q)}$, i.e. matrix $\hat{\kap}_{21}$ is obtained by the Hermitian conjugation of the coupling matrix $\hat{\kap}_{12}$:
\begin{equation}\label{Kappa21}
    \hat{\kap}_{21}=
    \begin{pmatrix}
    \kap & -i\Delta\\
    -i\Delta & -\gamma
    \end{pmatrix}
    \end{equation}
Using full-wave numerical simulations in CST Microwave Studio software package, we evaluate the parameters of the coupling matrices corresponding to the simulations Fig.~2 of the article main text as follows: 
$\kap=0.0265/a_z$, $\gamma=0.0118/a_z$, $\Delta=0.0263/a_z$, where $a_z$ is the period of the waveguide.

%\red{$\kap/k_{z0}=0.011$, $\gamma/k_{z0}=0.0049$, $\Delta/k_{z0}=0.0109$, where $k_{z0}$ is the Bloch wavenumber of an isolated waveguide at the frequency where two modes are degenerate.}

\section{Zak phase calculation}

Topological properties of the designed array of two-mode waveguides were examined in the main text using an extended unit cell for winding number calculation. Here we provide more direct though less intuitive derivation based on the direct calculation of the Zak phase.

Periodic part of the wave function corresponding to the lowest band of the Bloch Hamiltonian Eq.~(4) of the article main text is determined from the equations
\begin{equation}\label{EigEqAppD}
    \begin{pmatrix}
    (\kap+\gamma)\,\cos k+r & -2\Delta\,\sin k\\
    -2\Delta\,\sin k & -(\kap+\gamma)\,\cos k+r
    \end{pmatrix}
    \,
    \begin{pmatrix}
    u\\
    v
    \end{pmatrix}
    =0\:,
\end{equation}
where $r=\sqrt{(\kap+\gamma)^2\,\cos^2 k+4\Delta^2\,\sin^2 k}$. Accordingly,
\begin{equation}
\begin{split}
    & u=2N\,\Delta\,\sin k\:,\\
    & v=N\,\left[(\kap+\gamma)\,\cos k+r\right]\:,\\
    & N=\left[2r\,\left(r+(\kap+\gamma)\,\cos k\right)\right]^{-1/2}\:.
\end{split}
\end{equation}
Note that with these conventions both $u$ and $v$ are purely real. Furthermore, $v$ is positive for all $k$, while $u$ changes its sign. In the limiting case $k\rightarrow 0$ $(u,v)\rightarrow (0,1)$, and if  $k\rightarrow \pi$ $(u,v)\rightarrow (1,0)$.

The Zak phase is defined as
\begin{equation}\label{ZakPhaseDef}
    \gamma=i\,\int\limits_0^{2\pi}\,(u^*,v^*)\,
    \begin{pmatrix}
    \partial u/\partial k\\
    \partial v/\partial k
    \end{pmatrix}\,dk\:.
\end{equation}
Next we present $u=|u|\,e^{i\alpha}$. Taking into account the normalization condition $|u|^2+v^2=1$, we obtain
\begin{equation}\label{ZakPhase2}
    \gamma=-\int\limits_0^{2\pi}\,|u|^2\,\df{\alpha}{k}\,dk\:.
\end{equation}
The phase of $u$ changes from $\pi$ to $0$ at $k=0$ and changes from $0$ to $\pi$ at $k=\pi$, similar singularities are also observed at other points with $k=n\,\pi$. Therefore,
\begin{equation}
    \df{\alpha}{k}=-\pi\,\sum\limits_n \delta(k-2n\,\pi)+\pi\,\sum\limits_n \delta(k-(2n+1)\,\pi)\:.
\end{equation}
Since $u(2n\pi)\rightarrow 0$, the only point which gives the contribution to the integral Eq.~\eqref{ZakPhase2} is $k=\pi$. Thus, we finally get $\gamma=-\pi$ which indicates that the studied photonic band is topologically nontrivial in agreement with the reasoning in the main text.

\section{Detailed analysis of the edge modes in numerical simulations}\label{sec:AppE}

% %%%%%%%%%%%%%%%%%%%%%%%%%%%%%%%%%
%   \begin{center}
%     \begin{figure}[h]
%     \includegraphics[width=1.0\linewidth]{Fields_1D_CST.pdf}
%     \caption{Typical dependencies of the components $E_y$, $H_x$ and $H_z$ of electromagnetic field of the edge mode on $x$ coordinate. (a) Amplitudes. (b) Phases. \red{arg in vertical axis label on the bottom panel shouldn't be in italic.}} 
%     \label{fig:appendix_field_profiles}
%     \end{figure}
%     \end{center}
% %%%%%%%%%%%%%%%%%%%%%%%%%%%%%%%%%

%%%%%%%%%%%%%%%%%%%%%%%%%%%%%%%%%
   \begin{center}
    \begin{figure}[h]
    \includegraphics[width=1.0\linewidth]{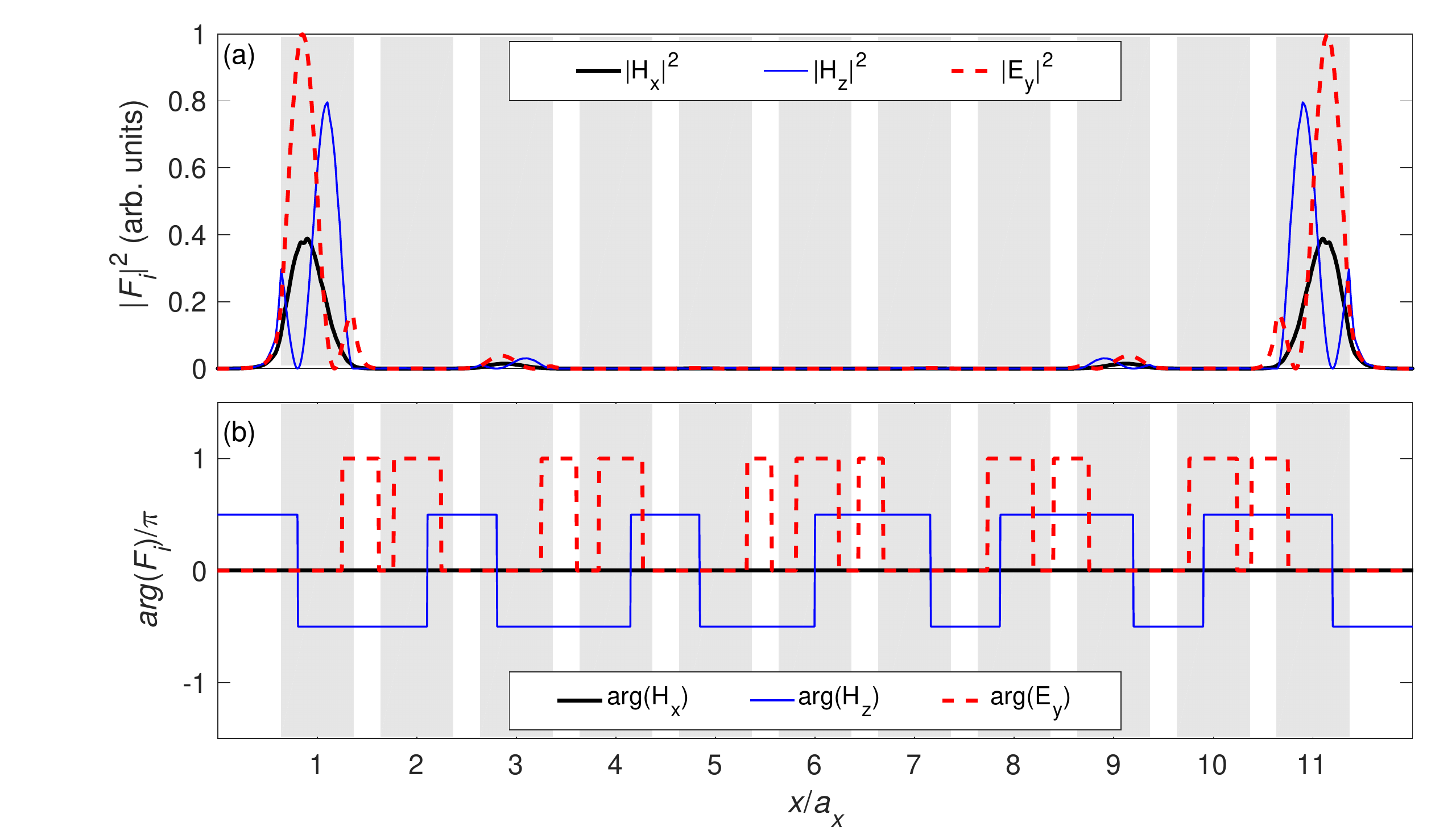}
    \caption{Typical dependencies of the components $E_y$, $H_x$ and $H_z$ of the edge mode field on $x$ coordinate along $x$ axis: (a) squared amplitudes, (b) phase profiles for one of the edge modes. Magnetic field is symmetric with respect to the reflection in $x=6$ plane.} 
    \label{fig:appendix_field_profiles}
    \end{figure}
    \end{center}
%%%%%%%%%%%%%%%%%%%%%%%%%%%%%%%%%

%%%%%%%%%%%%%%%%%%%%%%%%%%%%%%%%%
   \begin{center}
    \begin{figure}[h]
    \includegraphics[width=1.0\linewidth]{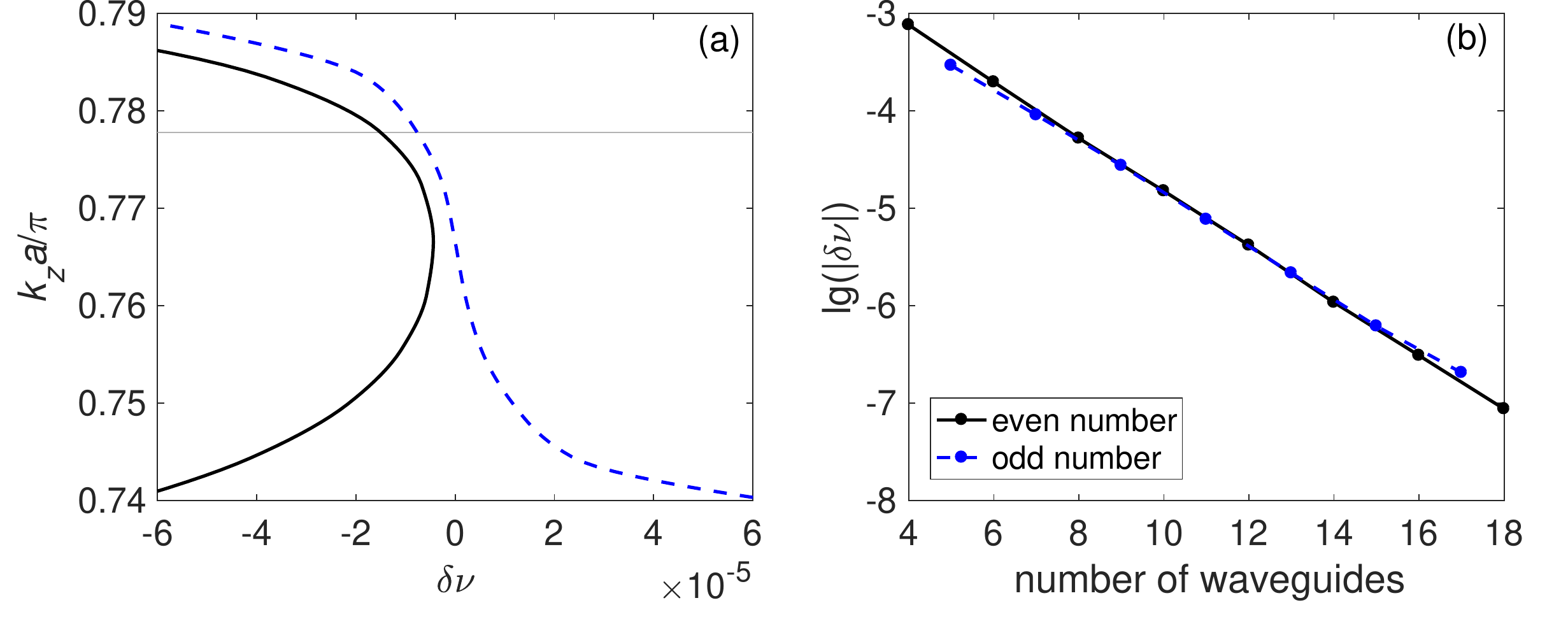}
    \caption{(a) Normalized eigenfrequency detuning $\delta\nu = (\nu_{edge_1} - \nu_{edge_2})/[(\nu_{edge_1} + \nu_{edge_2})/2]$ between two edge modes as a function of the propagation constant $k_z$ for $N=10$ (solid black curve) and $N=11$ (dashed blue curve) waveguides in the array. In the case of odd number of waveguides an accidental degeneracy is possible, whereas for the even number of waveguides avoided crossing is observed. (b) Dependence of  $\log\delta\nu$ on the number of waveguides for a fixed $k_z$ marked with horizontal grey line in panel (a). Exponential decrease is observed.} 
    \label{fig:appendix_delta_omega}
    \end{figure}
    \end{center}
%%%%%%%%%%%%%%%%%%%%%%%%%%%%%%%%%

In this Section, we present the detailed analysis of the edge eigenmodes in the considered system. Figure~\ref{fig:appendix_field_profiles} shows a typical distribution of amplitudes and phases for the dominant $E_y$, $H_x$ and $H_z$ components of electromagnetic field for one of the quasi-degenerate edge modes along the $x$ axis ($y=z=0$). Magnetic field of the considered mode is symmetric relative to the reflection in $x=6$ plane. Since the phase difference between $H_z$ and $H_x$ is equal to $+\pi/2$ ($-\pi/2$) inside the right (left) edge waveguide, magnetic field rotates around $y$ axis predominantly clockwise (counterclockwise)~\cite{Savelev-JAP}. This type of polarization is preserved for the arbitrary number of waveguides in the array.\par

Polarization state of the edge modes can be understood as follows. If the array is semi-infinite, it supports a single edge mode with the polarization handedness that depends on whether the edge waveguide is utmost left or utmost right one with respect to the propagation direction (positive direction of $z$ axis). In the case of a  finite array, there are two edge modes. If the number of waveguides is even, polarization of the fields at both edges is the same, and therefore two edge-localized field distributions interact with each other via their evanescent tails. As a consequence, there is an avoided crossing between symmetric and antisymmetric modes for any finite $N$~[Fig.~\ref{fig:appendix_delta_omega}(a)]. On the contrary, if the number of waveguides is odd, polarizations of the fields at both edges are different. Therefore, if the modes of a single waveguide are degenerate, the edge modes will be degenerate too, as illustrated in Fig.~\ref{fig:appendix_delta_omega}(a).

The frequency splitting between even and odd edge modes, however, can be neglected when the number of waveguides is large enough. This is illustrated by Fig.~\ref{fig:appendix_delta_omega}(b) showing the detuning $\delta\nu = 2\,(\nu_{edge_1} - \nu_{edge_2})/(\nu_{edge_1} + \nu_{edge_2})$ calculated for a fixed value of $k_z$ as a function of the number of the waveguides $N$.

\bibliography{TopologicalLib}